\documentclass[aps,twocolumn,showpacs,floatfix]{revtex4}
\usepackage{amsmath}
\usepackage{amsfonts}
\usepackage{amssymb}
\usepackage{epsfig}
\usepackage{graphicx}

\begin{document}
\title{Squeezing of a nanomechanical oscillator}
\author{Sumei Huang and G. S. Agarwal}
\affiliation{Department of Physics, Oklahoma State University,
Stillwater, Oklahoma 74078, USA}
\date{\today}

\begin{abstract}
We show that squeezing of a nanomechanical mirror can be generated
by injecting broad band squeezed vacuum light and laser light into
the cavity. We work in the resolved sideband regime. We find that in
order to obtain the maximum momentum squeezing of the movable
mirror, the squeezing parameter of the input light should be about
1. We can obtain more than $70\%$ squeezing. Besides, for a fixed
squeezing parameter, decreasing the temperature of the environment
or increasing the laser power increases the momentum squeezing. We
find very large squeezing with respect to thermal fluctuations, for
instance at 1 mK, the momentum fluctuations go down by a factor more
than one hundred.
\end{abstract}
 \maketitle

\renewcommand{\thesection}{\Roman{section}}
\setcounter{section}{0}
\section{Introduction}
\renewcommand{\baselinestretch}{1}\small\normalsize
    The optomechanical system has attracted much attention because of its potential
    applications in high precision measurements and quantum information
    processing ~\cite{LaHaye,Braginsky,Courty,Arcizet,Marshall,Caves,Pinard,Vitali1,Paternostro}.
Meanwhile, it provides a means of probing quantum behavior of a
macroscopic object if a nanomechanical oscillator can be cooled down
to near its quantum ground state ~\cite{Bhattacharya,Plenio}. Many
of these applications are becoming possible due to advances in
cooling the mirror
~\cite{Thompson,Bhattacharya3,Metzger,Gigan,Bouwmeester,Naik,Cohadon}.
Further as pointed out in Refs~\cite{Rae,Marquardt1,Schliesser}, the
ground state cooling can be achieved in the resolved sideband regime
where the frequency of the mechanical mirror is much larger than the
cavity decay rate.

Squeezing of a nanomechanical oscillator plays a vital role in
high-sensitive detection of position and force due to its less noise
in one quadrature than the coherent state. A number of different
methods have been developed to generate and enhance squeezing of a
nanomechanical oscillator, such as coupling a nanomechanical
oscillator to an atomic gas ~\cite{Ian}, a Cooper pair box
~\cite{Rabl}, a SQUID device ~\cite{Zhou}, using three-wave mixing
~\cite{Huo} or Circuit QED ~\cite{Moon}, or by means of quantum
measurement and feedback schemes
~\cite{Clerk,Woolley,Ruskov,Vitali}. A recent paper ~\cite{Jaehne}
reports squeezed state of a mechanical mirror can be created by
transfer of squeezing from a squeezed vacuum to a membrane within an
optical cavity under the conditions of ground state cooling. We
previously considered the possibility of using an OPA inside the
cavity for changing the nature of the statistical fluctuations
~\cite{Huang}.

In this paper, we propose a scheme that is capable of generating
squeezing of the movable mirror by feeding broad band squeezed
vacuum light along with the laser light. The achieved squeezing of
the mirror depends on the temperature of the mirror, the laser
power, and degree of squeezing of the input light. One can obtain
squeezing which could be more than 70\%.

The paper is structured as follows. In Sec. II we describe the
model, give the quantum Langevin equations, and obtain the
steady-state mean values. In Sec. III we derive the stability
conditions, calculate the mean square fluctuations in position and
momentum of the movable mirror. In Sec. IV we analyze how the
momentum squeezing of the movable mirror is affected by the
squeezing parameter, the temperature of the environment, and the
laser power. We also compare the momentum fluctuations of the
movable mirror in the presence of the coupling to the cavity field
with that in the absence of the coupling to cavity field. We find
very large squeezing with respect to thermal fluctuations, for
instance at 1 mK, the momentum fluctuations go down by a factor more
than one hundred. Our predictions of squeezing are based on the
parameters used in a recent experiment on normal mode splitting in a
nanomechanical oscillator ~\cite{Aspelmeyer}.

\section{Model}
\begin{figure}[htp]
 \scalebox{0.4}{\includegraphics{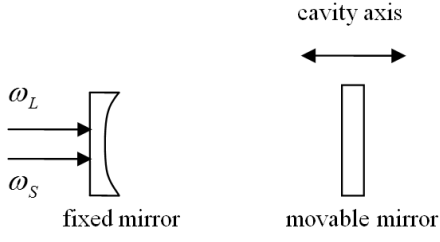}}
 \caption{\label{Fig1} Sketch of the studied system.
A laser with frequency $\omega_{L}$ and squeezed vacuum light with
frequency $\omega_{S}$ enter the cavity through the partially
transmitting mirror.}
 \end{figure}

The system to be considered, sketched in Fig.~\ref{Fig1}, is a
Fabry-Perot cavity with one fixed partially transmitting mirror and
one movable perfectly reflecting mirror in thermal equilibrium with
its environment at a low temperature. The cavity with length $L$ is
driven by a laser with frequency $\omega_{L}$, then the photons in
the cavity will exert a radiation pressure force on the movable
mirror due to momentum transfer. This force is proportional to the
instantaneous photon number in the cavity. The mirror also undergoes
thermal fluctuations due to environment. Under the effects of the
two forces, the movable mirror makes oscillation around its
equilibrium position. Here we treat the movable mirror as a quantum
mechanical harmonic oscillator with effective mass $m$, frequency
$\omega_{m}$ and momentum decay rate $\gamma_{m}$. We further assume
that the cavity is fed with squeezed light at frequency
$\omega_{S}$.

In the adiabatic limit, $\omega_{m}\ll\frac{c}{2L}$ ( $c$ is the
speed of light in vacuum), we ignore the scattering of photons to
other cavity modes, thus only one cavity mode $\omega_{c}$ is
considered ~\cite{Law}. In a frame rotating at the laser frequency,
the Hamiltonian for the system can be written as
\begin{eqnarray}\label{1}
H&=&\hbar(\omega_{c}-\omega_{L})n_{c}-\hbar g n_{c}
Q+\frac{\hbar\omega_{m}}{4}(Q^2+P^2)\nonumber\\
& &+i\hbar\varepsilon(c^{\dag}-c),
\end{eqnarray}
we have used the normalized coordinates for the oscillator defined
by $Q=\sqrt{\frac{2m\omega_{m}}{\hbar}}q$ and
$P=\sqrt{\frac{2}{m\hbar\omega_{m}}}p$ with $[Q,P]=2i$. This
normalization implies that in the ground state of the nanomechanical
mirror $\langle Q^{2}\rangle=\langle P^{2}\rangle=1$. Further in Eq.
(\ref{1}) the first term is the energy of the cavity field,
$n_{c}=c^{\dag}c$ is the number of the photons inside the cavity,
$c$ and $c^{\dag}$ are the annihilation and creation operators for
the cavity field with $[c,c^\dag]=1$. The second term comes from the
coupling of the movable mirror to the cavity field via radiation
pressure, the parameter
$g=\frac{\omega_{c}}{L}\sqrt{\frac{\hbar}{2m\omega_{m}}}$ is the
optomechanical coupling constant between the cavity and the movable
mirror. The third term corresponds the energy of the movable mirror.
The fourth term describes the coupling between the input laser field
and the cavity field, $\varepsilon$ is related to the input laser
power $\wp$ by $\varepsilon=\sqrt{\frac{2\kappa
\wp}{\hbar\omega_{L}}}$, where $\kappa$ is the cavity decay rate
associated with the transmission loss of the fixed mirror.

The equations of motion of the system can derived by the Heisenberg
equations of motion and adding the corresponding noise terms, this
gives the quantum Langevin equations
\begin{equation}\label{2}
\begin{array}{lcl}
\dot{Q}=\omega_{m}P,\vspace*{.1in}\\
\dot{P}=2g n_{c}-\omega_{m}Q-\gamma_{m}P+\xi,\vspace*{.1in}\\
\dot{c}=i(\omega_{L}-\omega_{c}+g
Q)c+\varepsilon-\kappa c+\sqrt{2
\kappa}c_{in},\vspace*{.1in}\\
\dot{c}^{\dag}=-i(\omega_{L}-\omega_{c}+g
Q)c^{\dag}+\varepsilon-\kappa
c^{\dag}+\sqrt{2\kappa}c_{in}^{\dag}.
\end{array}
\end{equation}
Here we have introduced the input squeezed vacuum noise operator
$c_{in}$ with frequency $\omega_S=\omega_L+\omega_m$. It has zero
mean value, and nonzero time-domain correlation
functions~\cite{Gardiner}
\begin{equation}\label{3}
\begin{array}{lcl}
\langle\delta c_{in}^{\dag}(t)\delta
c_{in}(t^{\prime})\rangle=N\delta(t-t^{\prime}),\vspace*{.1in}\\
\langle\delta c_{in}(t)\delta
c_{in}^{\dag}(t^{\prime})\rangle=(N+1)\delta(t-t^{\prime}),\vspace*{.1in}\\
\langle\delta c_{in}(t)\delta
c_{in}(t^{\prime})\rangle=Me^{-i\omega_{m}(t+t^{\prime})}\delta(t-t^{\prime}),\vspace*{.1in}\\
\langle\delta c_{in}^{\dag}(t)\delta
c_{in}^{\dag}(t^{\prime})\rangle=M^{\ast}e^{i\omega_{m}(t+t^{\prime})}\delta(t-t^{\prime}).
\end{array}
\end{equation}
where $N=\sinh^2(r)$, $M=\sinh(r)\cosh(r)e^{i\varphi}$, $r$ is the
squeezing parameter of the squeezed vacuum light, and $\varphi$ is
the phase of the squeezed vacuum light. For simplicity, we choose
$\varphi=0$. The force $\xi$ is the thermal Langevin force resulting
from the coupling of the movable mirror to the environment, whose
mean value is zero, and it has the following correlation function at
temperature $T$~\cite{Giovannetti}:
\begin{equation}\label{4}
\langle \xi(t)\xi(t^{'})\rangle=\frac{
\gamma_{m}}{\pi\omega_{m}}\int\omega e^{-i\omega (t-t^{'})}\left[1+\coth(\frac{\hbar
\omega}{2k_{B}T})\right]d\omega,
\end{equation}
where $k_B$ is the Boltzmann constant and $T$ is the temperature of
the environment. By using standard methods ~\cite{Walls}, setting
all the time derivatives in Eq. (\ref{2}) to zero, and solving it,
we obtain the steady-state mean values
\begin{equation}\label{5}
P_{s}=0,\hspace{.02in}Q_{s}=\frac{2g |c_{s}|^{2}}{\omega_{m}},\hspace{.02in}c_{s}=\frac{\varepsilon}{\kappa+i\Delta},
\end{equation}
where
\begin{equation}\label{6}
\Delta=\omega_{c}-\omega_{L}-g Q_{s}=\Delta_{0}-g
Q_{s}=\Delta_{0}-\frac{2g^2|c_{s}|^{2}}{\omega_{m}}
\end{equation}
is the effective cavity detuning, depending on $Q_{s}$. The $Q_{s}$
denotes the new equilibrium position of the movable mirror relative
to that without the driving field. Further $c_{s}$ represents the
steady-state amplitude of the cavity field. From Eq. (\ref{5}) and
Eq. (\ref{6}), we can see $Q_{s}$ satisfies a third order equation.
For a given detuning $\Delta_{0}$, $Q_{s}$ will at most have three
real values. Therefore, $Q_{s}$ and $c_{s}$ display an optical
multistable behavior ~\cite{Dorsel,Meystre,Marquardt}, which is a
nonlinear effect induced by the radiation-pressure coupling of the
movable mirror to the cavity field.

\section{Radiation pressure and quantum fluctuations}
To study squeezing of the movable mirror, we need to calculate the
fluctuations in the mirror's amplitude. Assuming that the nonlinear
coupling between the cavity field and the movable mirror is weak,
the fluctuation of each operator is much smaller than the
corresponding steady-state mean value, thus we can linearize the
system around the steady state. Writing each operator of the system
as the sum of its steady-state mean value and a small fluctuation
with zero mean value,
\begin{equation}\label{7}
Q=Q_{s}+\delta Q,\hspace*{.1in}P=P_{s}+\delta P,\hspace*{.1in}c=c_{s}+\delta c.
\end{equation}
Inserting Eq. (\ref{7}) into Eq. (\ref{2}), then assuming
$|c_{s}|\gg1$, the linearized quantum Langevin equations for the
fluctuation operators can be expressed as follows,
\begin{equation}\label{8}
\begin{array}{lcl}
\delta\dot{Q}=\omega_{m}\delta P,\vspace*{.1in}\\
\delta\dot{P}=2g (c^\ast_{s}\delta c+c_{s}\delta
c^{\dag})-\omega_{m}\delta Q-\gamma_{m}\delta P+\xi,\vspace{.1in}\\
\delta\dot{c}=-(\kappa+i\Delta)\delta c+ig c_{s}\delta
Q+\sqrt{2 \kappa}\delta c_{in},\vspace*{.1in}\\
\delta\dot{c}^{\dag}=-(\kappa-i\Delta)\delta c^{\dag}-ig
c^{*}_{s}\delta Q+\sqrt{2 \kappa}\delta c_{in}^{\dag}.
\end{array}
\end{equation}
Introducing the cavity field quadratures $\delta x=\delta c+\delta
c^{\dag}$ and $\delta y=i(\delta c^{\dag}-\delta c)$, and the input
noise quadratures $\delta x_{in}=\delta c_{in}+\delta c_{in}^{\dag}$
and $\delta y_{in}=i(\delta c_{in}^{\dag}-\delta c_{in})$, Eq.
(\ref{8}) can be rewritten in the matrix form
\begin{equation}\label{9}
\dot{f}(t)=Af(t)+\eta(t),
\end{equation}
in which $f(t)$ is the column vector of the fluctuations, $\eta(t)$
is the column vector of the noise sources. Their transposes are
\begin{equation}\label{10}
\begin{array}{lcl}
f(t)^{T}=(\delta Q,\delta P,\delta x,\delta y),\vspace*{.1in}\\
\eta(t)^{T}=(0,\xi,\sqrt{2\kappa}\delta
x_{in},\sqrt{2\kappa}\delta y_{in});
\end{array}
\end{equation}
and the matrix $A$ is given by
\begin{equation}\label{11}
A=\left(
  \begin{array}{cccc}
    0&\omega_{m} & 0 & 0 \\
    -\omega_{m}&-\gamma_{m} & g (c_{s}+c_{s}^{*})& -ig(c_{s}-c_{s}^{*}) \\
    ig (c_{s}-c_{s}^{*})  & 0 & -\kappa & \Delta \\
     g (c_{s}+c_{s}^{*})  & 0 & -\Delta & -\kappa \\
  \end{array}
\right).
\end{equation}
The system is stable only if the real parts of all the eigenvalues of the matrix $A$
are negative. The stability conditions for the system
can be derived by applying the Routh-Hurwitz criterion
~\cite{Hurwitz,DeJesus}, we get
\begin{equation}\label{12}
\begin{array}{lcl}
\kappa\gamma_{m}[(\kappa^2+\Delta^2)^2+(2\kappa\gamma_{m}+\gamma_{m}^2-2\omega_{m}^2)(\kappa^2+\Delta^2)\vspace{.1in}\\
\hspace{.25in}+\omega_{m}^2(4\kappa^2+\omega_{m}^2+2\kappa\gamma_{m})]+2\omega_{m}\Delta g^2|c_{s}|^2\vspace{.1in}\\
\hspace{.25in}\times(2\kappa+\gamma_{m})^2>0,\vspace{.1in}\\
\omega_{m}(\kappa^{2}+\Delta^2)-4\Delta g^2|c_{s}|^2>0.
\end{array}
\end{equation}
All the external parameters chosen in this paper satisfy the
stability conditions (\ref{12}) to ensure the system to be stable.

Fourier transforming each operator in Eq. (\ref{8}) and solving it
in the frequency domain, the position fluctuations of the movable
mirror are given by
\begin{equation}\label{13}
\begin{array}{lcl}
\delta Q(\omega)=\frac{1}{d(\omega)}(2\sqrt{2\kappa}\omega_m g\{[\kappa-i(\Delta+\omega)]c_{s}^{*}\delta c_{in}(\omega)\vspace{.1in}\\
\hspace{.5in}+[\kappa+i(\Delta-\omega)]c_{s}\delta c_{in}^{\dag}(-\omega)\}\vspace{.1in}\\
\hspace{.5in}+\omega_m[(\kappa-i\omega)^2+\Delta^2]\xi(\omega)),
\end{array}
\end{equation}
where $d(\omega)=-4\omega_{m}\Delta
g^2|c_{s}|^2+(\omega_m^2-\omega^2-i\gamma_{m}\omega
)[(\kappa-i\omega)^2+\Delta^2]$. In Eq. (\ref{13}), the first term
proportional to $g$ originates from radiation pressure, while the
second term involving $\xi$ is from the thermal noise. So the
position fluctuations of the movable mirror are now determined by
radiation pressure and the thermal noise. In the case of no coupling
with the cavity field, the movable mirror will make Brownian motion,
$\delta
Q(\omega)=\omega_{m}\xi(\omega)/(\omega_{m}^2-\omega^2-i\gamma_{m}\omega)$,
whose susceptibility has a Lorentzian shape centered at frequency
$\omega_{m}$ with width $\gamma_{m}$.

 Taking Fourier transform of
$\delta\dot{Q}=\omega_{m}\delta P$ in Eq. (8), we further obtain the
momentum fluctuations of the movable mirror, $\delta
P(\omega)=-i\frac{\omega}{\omega_{m}}\delta Q(\omega)$.

The mean square fluctuations in position and momentum of the movable
mirror are determined by
\begin{equation}\label{14}
\begin{array}{lcl}
\langle\delta
Q(t)^{2}\rangle=\frac{1}{4\pi^{2}}\int\int_{-\infty}^{+\infty}
d\omega d\Omega e^{-i(\omega+\Omega)t} \langle\delta Q(\omega)\delta
Q(\Omega)\rangle,\vspace{.1in}\\
\langle\delta
P(t)^{2}\rangle=\frac{1}{4\pi^{2}}\int\int_{-\infty}^{+\infty}
d\omega d\Omega e^{-i(\omega+\Omega)t} \langle\delta P(\omega)\delta
P(\Omega)\rangle.
\end{array}
\end{equation}

To calculate the mean square fluctuations, we require the
correlation functions of the noise sources in the frequency domain,
\begin{equation}\label{15}
\begin{array}{lcl}
\langle\delta c_{in}^{\dag}(\omega)\delta
c_{in}(\Omega)\rangle=2\pi N\delta(\omega+\Omega),\vspace{.1in}\\
\langle\delta c_{in}(\omega)\delta
c_{in}^{\dag}(\Omega)\rangle=2\pi(N+1)\delta(\omega+\Omega),\vspace{.1in}\\
\langle\delta c_{in}(\omega)\delta
c_{in}(\Omega)\rangle=2\pi M\delta(\omega+\Omega-2\omega_{m}),\vspace{.1in}\\
\langle\delta c_{in}^{\dag}(\omega)\delta
c_{in}^{\dag}(\Omega)\rangle=2\pi M^{\ast}\delta(\omega+\Omega+2\omega_{m}),\vspace{.1in}\\
\langle\xi(\omega)\xi(\Omega)\rangle=4\pi\gamma_{m}\frac{\omega
}{\omega_{m}}\left[1+\coth(\frac{\hbar\omega}{2k_B T})\right]\delta(\omega+\Omega).
\end{array}
\end{equation}
Combining Eqs. (\ref{13}) -- (\ref{15}), after some calculations,
the mean square fluctuations of  Eq. (\ref{14}) are written as
\begin{equation}\label{16}
\begin{array}{lcl}
\langle\delta Q(t)^{2}\rangle=\frac{1}{2\pi}\int_{-\infty}^{+\infty}
\omega_m^2(A+B e^{-2i\omega_{m}t}+C e^{2i\omega_{m}t})
d\omega,\vspace{.1in}\\ \langle\delta
P(t)^{2}\rangle=\frac{1}{2\pi}\int_{-\infty}^{+\infty} [\omega^2
A+\omega (\omega-2\omega_m) B
e^{-2i\omega_{m}t}\vspace{.1in}\\\hspace{.7in}+\omega
(\omega+2\omega_m) C e^{2i\omega_{m}t}] d\omega.
\end{array}
\end{equation}
where \begin{equation}\label{17}
\begin{array}{lcl}A=\frac{1}{d(\omega)d(-\omega)}(8\kappa
g^2|c_{s}|^2\{(N+1)[\kappa^2+(\Delta+\omega)^2]\vspace{.1in}\\\hspace{.2in}
+N[\kappa^2+(\Delta-\omega)^2]\}+2\gamma_{m}\frac{\omega}{\omega_{m}}[(\Delta^2+\kappa^2-\omega^2)^2\vspace{.1in}\\\hspace{.2in}+4\kappa^2\omega^2][1+\coth(\frac{\hbar\omega}{2k_B
T})]), \vspace{.1in}\\
B=\frac{8\kappa g^2 c_{s}^{*2}M}{d(\omega)d(2\omega_{m}-\omega)}
[\kappa-i(\Delta+\omega)][\kappa-i(\Delta+2\omega_{m}-\omega)],\vspace{.1in}\\
C=\frac{8\kappa g^2
c_{s}^{2}M^{*}}{d(\omega)d(-2\omega_{m}-\omega)}[\kappa+i(\Delta-\omega)][\kappa+i(\Delta+2\omega_{m}+\omega)].\vspace{.1in}\\
\end{array}
\end{equation}
In Eqs. (\ref{16}) and (\ref{17}), the term independent of $g$ is
from the thermal noise contribution; while those terms involving $g$
arise from the radiation pressure contribution, including the
influence of the squeezed vacuum light. Moreover, either
$\langle\delta Q(t)^{2}\rangle$ or $\langle\delta P(t)^{2}\rangle$
contains three terms, the first term is independent of time, but the
second and third terms are time-dependent, which causes
$\langle\delta Q(t)^{2}\rangle$ and $\langle\delta P(t)^{2}\rangle$
vary with time. The complex exponential in Eq. (\ref{16}) can be
removed by working in the interaction picture. Let's define $b$
($b^\dag$) and $\tilde{b}$ ($\tilde{b}^\dag$) be the annihilation
(creation) operators for the oscillator in the Schr\"{o}dinger and
interaction picture with $[b,b^\dag]=1$ and
$[\tilde{b},\tilde{b}^\dag]=1$. The relations between them are
$b=\tilde{b} e^{-i\omega_{m}t}$ and $b^\dag=\tilde{b}^\dag
e^{i\omega_{m}t}$. Then using $Q=b+b^\dag$, $P=i(b^\dag-b)$,
 $\tilde{Q}=\tilde{b}+\tilde{b}^\dag$, and
$\tilde{P}=i(\tilde{b}^\dag-\tilde{b})$, we get
\begin{equation}\label{18}
\begin{array}{lcl}
\langle\delta
\tilde{Q}^{2}\rangle=\frac{1}{2\pi}\int_{-\infty}^{+\infty}
\omega_m^2(A+B+C) d\omega,\vspace{.1in}\\
\langle\delta
\tilde{P}^{2}\rangle=\frac{1}{2\pi}\int_{-\infty}^{+\infty}
[\omega^2 A+\omega (\omega-2\omega_m)
B\vspace{.1in}\\\hspace{.5in}+\omega (\omega+2\omega_m) C] d\omega.
\end{array}
\end{equation}

 According to the Heisenberg uncertainty
principle,
\begin{equation}\label{19}
\langle\delta \tilde{Q}^{2}\rangle\langle\delta
\tilde{P}^{2}\rangle\ge|\frac{1}{2}[\tilde{Q},\tilde{P}]|^{2}.
\end{equation}
If either $\langle\delta \tilde{Q}^{2}\rangle<1$ or $\langle\delta
\tilde{P}^{2}\rangle<1$, the movable mirror is said to be squeezed.

From Eqs. (\ref{17}) and (\ref{18}), we find $\langle\delta
\tilde{Q}^{2}\rangle$ or $\langle\delta \tilde{P}^{2}\rangle$ is
determined by the detuning $\Delta_0$, the squeezing parameter $r$,
the laser power $\wp$, the cavity length $L$, the temperature of the
environment $T$, and so on. Here we focus on the dependence of
$\langle\delta \tilde{Q}^{2}\rangle$ and $\langle\delta
\tilde{P}^{2}\rangle$ on the squeezing parameter, the temperature of
the environment, and the laser power.

\section{Squeezing of the movable mirror}
In this section, we numerically evaluate the mean square
fluctuations in position and momentum of the movable mirror given by
Eq. (\ref{18}) to show squeezing of the movable mirror produced by
feeding the squeezed vacuum light at the input mirror. We use the
same parameters as those in the recent successful experiment on
normal mode splitting in a nanomechanical oscillator
~\cite{Aspelmeyer}: the wave length of the laser $\lambda=\frac{2\pi
c}{\omega_L}=1064$ nm, $L=25$ mm, $m=145$ ng,
$\kappa=2\pi\times215\times10^3$ Hz,
$\omega_m=2\pi\times947\times10^3$ Hz, the mechanical quality factor
$Q^{\prime}=\frac{\omega_{m}}{\gamma_{m}}=6700$. In the case of $k_B
T\gg\hbar\omega_{m}$, we may approximate $\coth(\hbar\omega/(2k_B
T))\simeq2k_B T/(\hbar\omega)$. In the case of $T=0$ K, if
$\omega<0$, $ \coth(\hbar\omega/(2k_B T))\simeq-1$, if $\omega>0$,
$\coth(\hbar\omega/(2k_B T))\simeq1$. Through numerical
calculations, it is found that squeezing of $\langle\delta
\tilde{Q}^{2}\rangle$ doesn't exist but squeezing of $\langle\delta
\tilde{P}^{2}\rangle$ exists. In the following we therefore
concentrate on discussing $\langle\delta \tilde{P}^{2}\rangle$.

Note that in the absence of the coupling to the cavity field, the
movable mirror is in free space, and is coupled to the environment.
Then the fluctuations are given by
\begin{eqnarray}\label{20}
\langle\delta \tilde{Q}^{2}\rangle=\langle\delta
\tilde{P}^{2}\rangle&=&1+\frac{2}{e^{\hbar \omega_m/(k_B T)}-1}\nonumber\\
&=& \left\{\begin{array}{ll}
 1 & \mbox{for $T=0$ K},\\
 44 &\mbox{for $T=1$ mK},\\
 440 &\mbox{for $T=10$ mK}.
 \end{array}
 \right.
 \end{eqnarray}
As well known no squeezing of the movable mirror occurs.

\begin{figure}[htp]
 \scalebox{0.65}{\includegraphics{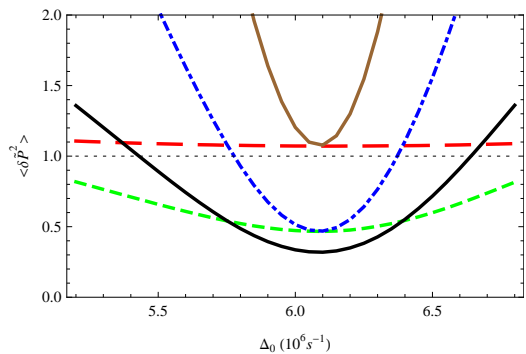}}
 \caption{\label{Fig2}(Color online)  The mean square fluctuations $\langle\delta \tilde{P}^{2}\rangle$ versus the detuning $\Delta_0$ ($10^6$ s$^{-1}$) for different values of the squeezing of the input field. $r=0$ (red, big dashed line), $r=0.5$ (green, small dashed line), $r=1$ (black, solid curve), $r=1.5$ (blue, dotdashed curve), $r=2$ (brown, solid curve). The minimum values of $\langle\delta \tilde{P}^{2}\rangle$ are 1.071 ($r$=0), 0.467 ($r$=0.5), 0.319 ($r$=1),
0.468 ($r$=1.5), 1.078 ($r$=2). The flat dotted line represents the
variance of the coherent light ($\langle\delta
\tilde{P}^{2}\rangle$=1). Parameters: the temperature of the
environment $T=1$ mK, the laser power $\wp=6.9$ mW.}
 \end{figure}

Now we consider fluctuations in the presence of the coupling to the
cavity field. If we choose $T=1$ mK, and $\wp=6.9$ mW, the mean
square fluctuations $\langle\delta \tilde{P}^{2}\rangle$ are plotted
as a function of the detuning $\Delta_{0}$ in the Fig.~\ref{Fig2}.
Different graphs correspond to different values of the squeezing of
the input light.  In the case of no injection of the squeezed vacuum
light ($r=0$), which means that the squeezed vacuum light is
replaced by an ordinary vacuum light, we find $\langle\delta
\tilde{P}^{2}\rangle$ is always larger than unity (the coherent
level), the minimum value of $\langle\delta \tilde{P}^{2}\rangle$ is
1.071, thus there is no momentum squeezing of the movable mirror.
However, if we inject the squeezed vacuum light, it is seen that the
momentum squeezing of the movable mirror occurs, and the maximum
squeezing happens at about $r=1$, the corresponding minimum value of
$\langle\delta \tilde{P}^{2}\rangle$ is 0.319, thus the maximum
amount of squeezing is about $68\%$.  So the injection of the
squeezed vacuum light greatly reduces the fluctuations in momentum,
because using the squeezed vacuum light increases the photon number
in the cavity, which results in a stronger radiation pressure acting
on the movable mirror. Note that the minimum value of $\langle\delta
\tilde{P}^{2}\rangle$ in the presence of the coupling to the cavity
field is much less than that ($\langle\delta
\tilde{P}^{2}\rangle=44$) in the absence of the coupling to the
cavity field. So there is very large squeezing with respect to
thermal fluctuations. The momentum fluctuations can be reduced by a
factor more than one hundred.

\begin{figure}[htp]
 \scalebox{0.65}{\includegraphics{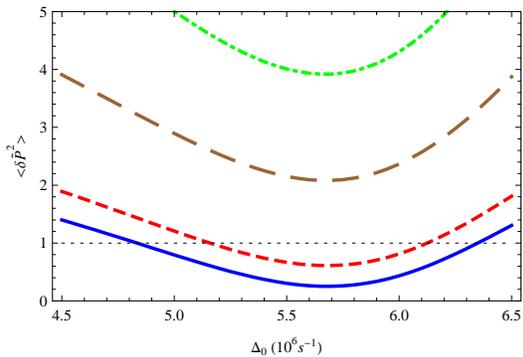}}
 \caption{\label{Fig3}(Color online)  The mean square fluctuations $\langle\delta \tilde{P}^{2}\rangle$ versus the detuning $\Delta_0$ ($10^6$ s$^{-1}$), each curve corresponds to a
different temperature of the environment. $T$=0 K (blue, solid
curve), 1 mK (red, small dashed curve), 5 mK (brown, big dashed
curve), 10 mK (green, dotdashed curve). The minimum values of
$\langle\delta \tilde{P}^{2}\rangle$ are 0.252 ($T$=0 K), 0.611
($T$=1 mK), 2.082 ($T$=5 mK), 3.919 ($T$=10 mK). The flat dotted
line represents the variance of the coherent light ($\langle\delta
\tilde{P}^{2}\rangle$=1). Parameters: the squeezing parameter $r=1$,
the laser power $ \wp=0.6$ mW.}
 \end{figure}

\begin{figure}[htp]
 \scalebox{0.65}{\includegraphics{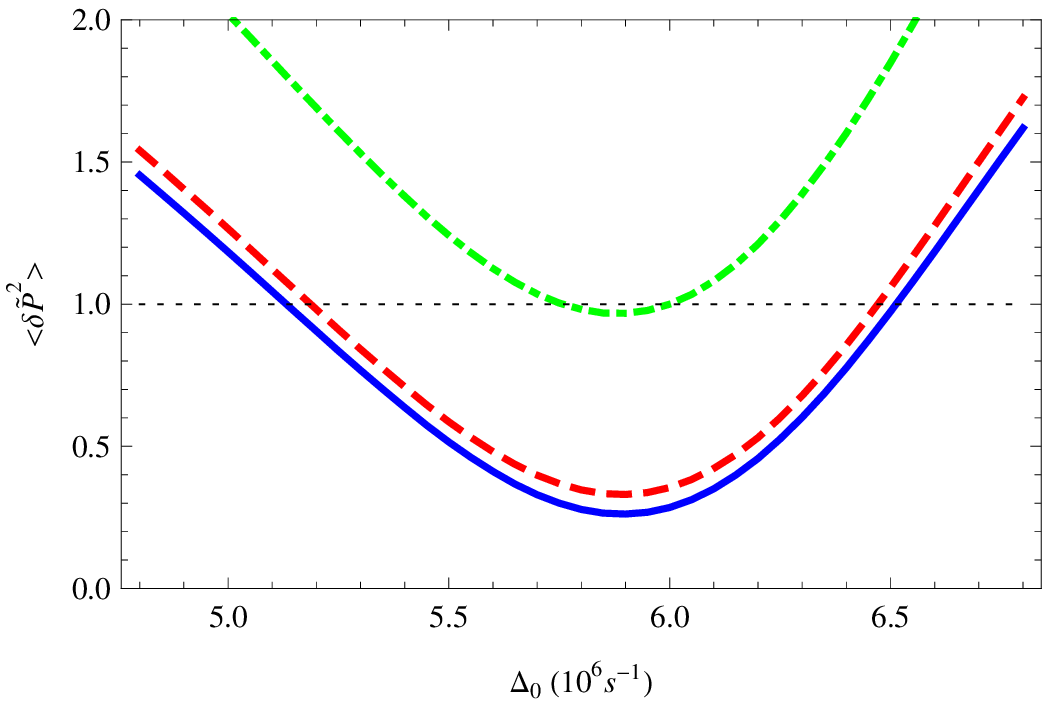}}
 \caption{\label{Fig4}(Color online)   The mean square fluctuations $\langle\delta \tilde{P}^{2}\rangle$ versus  the detuning $\Delta_0$ ($10^6$ s$^{-1}$), each curve corresponds to a
different temperature of the environment. $T$=0 K (solid curve), 1
mK (dashed curve), 10 mK (dotdashed curve). The minimum values of
$\langle\delta \tilde{P}^{2}\rangle$ are 0.261 ($T$=0 K), 0.330
($T$=1 mK), 0.968 ($T$=10 mK). The flat dotted line represents the
variance of the coherent light ($\langle\delta
\tilde{P}^{2}\rangle$=1). Parameters: the squeezing parameter $r=1$,
the laser power $\wp=3.8$ mW.}
\end{figure}
\begin{figure}[htp]
 \scalebox{0.65}{\includegraphics{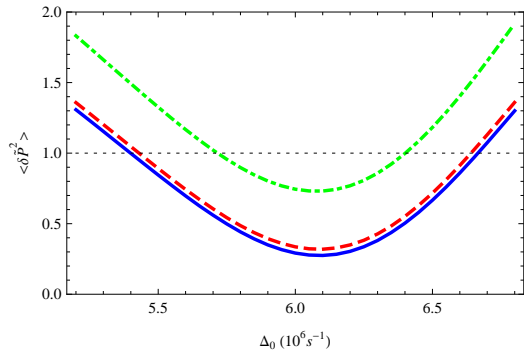}}
 \caption{\label{Fig5}(Color online)  The mean square fluctuations $\langle\delta \tilde{P}^{2}\rangle$ versus the detuning $\Delta_0$ ($10^6$ s$^{-1}$), each curve corresponds to a
different temperature of the environment. $T$=0 K (solid curve), 1
mK (dashed curve), 10 mK (dotdashed curve). The minimum values of
$\langle\delta \tilde{P}^{2}\rangle$ are 0.275 ($T$=0 K), 0.319
($T$=1 mK), 0.731 ($T$=10 mK). The flat dotted line represents the
variance of the coherent light ($\langle\delta
\tilde{P}^{2}\rangle$=1). Parameters: the squeezing parameter $r=1$,
the laser power $ \wp=6.9$ mW.}
 \end{figure}

Then we fix the squeezing parameter $r=1$, the mean square
fluctuations $\langle\delta \tilde{P}^{2}\rangle$ as a function of
the detuning $\Delta_0$ for different temperature of the environment
and laser power are shown in Figs.~\ref{Fig3} --~\ref{Fig5}. For a
given lase power, we find that the minimum value of $\langle\delta
\tilde{P}^{2}\rangle$ decreases with decrease of the temperature of
the environment as expected. The lower is the temperature, the less
is the thermal noise. At $T=0$ K, the minimum value of
$\langle\delta \tilde{P}^{2}\rangle$ is the smallest due to no
thermal noise, which corresponds to the maximum momentum squeezing
of the movable mirror. For example, when $T=0$ K and $\wp=0.6$ mW,
the minimum value of $\langle\delta \tilde{P}^{2}\rangle$ is 0.252,
the corresponding amount of squeezing is up to about $75\%$.
Therefore, decreasing the temperature of the environment can enhance
the amount of the momentum squeezing of the movable mirror. On the
other hand, we note that when the temperature of the environment is
high, for example, for $T=10$ mK, and laser power 0.6 mW, the
minimum value of $\langle\delta \tilde{P}^{2}\rangle$ is 3.919. In
this case, there is no momentum squeezing, but if we increase the
laser power to 6.9 mW, the minimum value of $\langle\delta
\tilde{P}^{2}\rangle$ is 0.731, the movable mirror shows momentum
squeezing, and the amount of squeezing will increase with increase
of laser power. Therefore, when the temperature of the environment
is high, the momentum squeezing of the movable mirror can be
obtained by increasing the input laser power. The reason is that
increasing the laser power can increase the photon number in the
cavity. Moreover, for any specific temperature of the environment,
the minimum value of $\langle\delta \tilde{P}^{2}\rangle$ in the
presence of the radiation pressure coupling is always much less than
that in the absence of the radiation pressure coupling.

\section{Conclusions}
In conclusion, we have found that squeezing of the movable mirror
can be achieved by the injection of squeezed vacuum light and a
laser. The result shows the maximum momentum squeezing of the
movable mirror happens if squeezed vacuum light with $r$ about 1 is
injected into the cavity. For a given squeezing parameter and laser
power, decreasing the temperature of the environment can enhance the
maximum momentum squeezing of the movable mirror. In addition, the
momentum squeezing of the movable mirror may be achieved by
increasing the input laser power. Generation of squeezing of the
movable mirror provides a new way to detect a weak force. Further
the ``feeding" of squeezed light can be used to squeeze collective
degrees of freedom for several mirrors inside the cavity.

GSA thanks Jeff Kimble for suggesting the possibility of generating
squeezing by pumping the cavity with squeezed vacuum light, instead
of using an OPA inside the cavity ~\cite{Huang}. We gratefully
acknowledge support for NSF Grants CCF 0829860 and Phys. 0653494.


\begin{thebibliography}{99}

\bibitem{LaHaye} M. D. LaHaye, O. Buu, B. Camarota, and K. C. Schwab, Science \textbf{304}, 74 (2004).
\bibitem{Braginsky} V. B. Braginsky and F. Y. Khalili, \textit{Quantum Measurement}
(Cambridge University Press, Cambridge, 1992).
\bibitem{Courty} J. M. Courty, A. Heidmann, and M. Pinard, Phys. Rev. Lett. \textbf{90}, 083601 (2003).
\bibitem{Arcizet} O. Arcizet, T. Briant, A. Heidmann, and M. Pinard, Phys. Rev. A \textbf{73}, 033819 (2006).
\bibitem{Marshall} W. Marshall, C. Simon, R. Penrose, and D. Bouwmeester, Phys. Rev. Lett. \textbf{91}, 130401 (2003).
\bibitem{Caves} C. M. Caves, Phys. Rev. Lett. \textbf{45}, 75 (1980).
\bibitem{Pinard} M. Pinard, A. Dantan, D. Vitali, O. Arcizet, T. Briant, and A. Heidmann,  Europhys. Lett. \textbf{72}, 747 (2005).
\bibitem{Vitali1} D. Vitali, S. Gigan, A. Ferreira, H. R. B\"{o}hm, P. Tombesi, A. Guerreiro, V. Vedral, A. Zeilinger, and M. Aspelmeyer, Phys. Rev. Lett. \textbf{98}, 030405 (2007).
\bibitem{Paternostro} M. Paternostro, D. Vitali, S. Gigan, M. S. Kim, C. Brukner, J. Eisert, and M. Aspelmeyer, Phys. Rev. Lett. \textbf{99}, 250401 (2007).
\bibitem{Bhattacharya} M. Bhattacharya, P.-L. Giscard, and P. Meystre, Phys. Rev. A \textbf{77}, 030303(R) (2008).
\bibitem{Plenio} M. J. Hartmann and M. B. Plenio, Phys. Rev. Lett. \textbf{101}, 200503
(2008).
\bibitem{Thompson} J. D. Thompson, B. M. Zwickl, A. M.
Jayich, F. Marquardt, S. M. Girvin, and J. G. E. Harris, Nature
(London) \textbf{452}, 72 (2008).
\bibitem{Bhattacharya3} M. Bhattacharya, H. Uys, and P. Meystre, Phys. Rev. A \textbf{77}, 033819 (2008).
\bibitem{Metzger}C. H. Metzger and K. Karrai, Nature (London) \textbf{432}, 1002 (2004).
\bibitem{Gigan} S. Gigan, H. R. B\"{o}hm, M. Paternostro, F. Blaser, G. Langer, J. B.
Hertzberg, K. C. Schwab, D. B\"{a}uerle, M. Aspelmeyer, and A.
Zeilinger, Nature (London) \textbf{444}, 67 (2006).
\bibitem{Bouwmeester} D. Kleckner and D. Bouwmeester, Nature (London) \textbf{444}, 75 (2006).
\bibitem{Naik} A. Naik, O. Buu, M. D. LaHaye, A. D. Armour, A. A. Clerk, M. P.
Blencowe, and K. C. Schwab, Nature (London) \textbf{443}, 193
(2006).
\bibitem{Cohadon} O. Arcizet, P.-F. Cohadon, T. Briant, M. Pinard, and A.
Heidmann, Nature (London) \textbf{444}, 71 (2006).
\bibitem{Rae} I. Wilson-Rae, N. Nooshi, W. Zwerger, and T. J. Kippenberg, Phys. Rev. Lett. \textbf{99}, 093901 (2007).
\bibitem{Marquardt1} F. Marquardt, J. P. Chen, A. A. Clerk, and S. M. Girvin, Phys. Rev. Lett. \textbf{99}, 093902 (2007).
\bibitem{Schliesser} A. Schliesser, R. Rivière, G. Anetsberger, O. Arcizet, and T. J. Kippenberg, Nature Physics \textbf{4}, 415 (2008).
\bibitem{Ian} H. Ian, Z. R. Gong, Y. X. Liu, C. P. Sun, and F. Nori, Phys. Rev. A \textbf{78}, 013824 (2008).
\bibitem{Rabl} P. Rabl, A. Shnirman, and P. Zoller, Phys. Rev. B \textbf{70}, 205304 (2004).
\bibitem{Zhou} X. Zhou and A. Mizel, Phys. Rev. Lett. \textbf{97}, 267201 (2006).
\bibitem{Huo} W. Huo and G. Long, Appl. Phys. Lett. \textbf{92}, 133102 (2008).
\bibitem{Moon} K. Moon and S. M. Girvin, Phys. Rev. Lett. \textbf{95}, 140504 (2005).
\bibitem{Clerk} A. A. Clerk, F. Marquardt, and K. Jacobs, New J. Phys. \textbf{10}, 095010 (2008).
\bibitem{Woolley} M. J. Woolley, A. C. Doherty, G. J. Milburn, and K. C. Schwab, Phys. Rev. A \textbf{78}, 062303 (2008).
\bibitem{Ruskov} R. Ruskov, K. Schwab, and A. N. Korotkov, Phys. Rev. B \textbf{71}, 235407 (2005).
\bibitem{Vitali} D. Vitali, S. Mancini, L. Ribichini, and P. Tombesi, Phys. Rev. A \textbf{65}, 063803 (2002).
\bibitem{Jaehne} K. J\"{a}ehne, C. Genes, K. Hammerer, M. Wallquist, E. S. Polzik, and P. Zoller, arXiv: quant-ph/0904.1306v1.
\bibitem{Huang} S. Huang and G. S. Agarwal, Phys. Rev. A \textbf{79}, 013821 (2009).
\bibitem{Aspelmeyer} S. Gr\"{o}blacher, K. Hammerer, M. R. Vanner, and M. Aspelmeyer, arXiv: quant-ph/0903.5293v1.
\bibitem{Law} C. K. Law, Phys. Rev. A \textbf{49}, 433 (1994); \textit{ibid.} \textbf{51}, 2537 (1995);
\bibitem{Gardiner} C. W. Gardiner, Phys. Rev. Lett. \textbf{56}, 1917 (1986).
\bibitem{Giovannetti} V. Giovannetti and D. Vitali, Phys. Rev. A \textbf{63}, 023812 (2001).
\bibitem{Walls} D. F. Walls and G. J. Milburn, \textit{Quantum Optics} (Springer-Verlag, Berlin, 1998).
\bibitem{Dorsel} A. Dorsel, J. D. McCullen, P. Meystre, E. Vignes, and H. Walther, Phys. Rev. Lett. \textbf{51}, 1550 (1983).
\bibitem{Meystre} P. Meystre, E. M. Wright, J. D. McCullen, and E. Vignes, J. Opt. Soc. Am. B \textbf{2}, 1830 (1985).
\bibitem{Marquardt} F. Marquardt, J. G. E. Harris, and S. M.
Girvin, Phys. Rev. Lett. \textbf{96}, 103901 (2006).
\bibitem{Hurwitz} A. Hurwitz, \textit{Selected Papers on
Mathematical  Trends in Control Theory}, edited by R. Bellman and R.
Kalaba (Dover, New York, 1964).
\bibitem{DeJesus} E. X. DeJesus and C. Kaufman, Phys. Rev. A \textbf{35}, 5288 (1987).


\end{thebibliography}
\end{document}